\documentclass[pra,aps,showpacs,groupedaddress,superscriptaddress,twocolumn,toc=flat]{revtex4-1}

\usepackage[colorlinks=true,linkcolor=blue,urlcolor=blue,citecolor=blue]{hyperref}

\usepackage[utf8x]{inputenc}
\usepackage{color}
\usepackage{bbm} 

\usepackage{amsfonts,amsmath,amssymb,stmaryrd}

\usepackage{graphicx}
\usepackage{subfigure}  % use for side-by-side figures
\usepackage{bbm} 
\usepackage{hyperref}
\usepackage{epsfig}
\usepackage{mathrsfs}
\usepackage{verbatim}
\usepackage{centernot}
\usepackage{ulem}
\usepackage{longtable}
\usepackage{nicefrac}
\usepackage[framemethod=tikz]{mdframed}
 
\renewcommand{\l}{\left(}
\renewcommand{\r}{\right)}

\newcommand{\Zt}{$\mathbb{Z}_2$~}

\newcommand{\bra}[1]{\langle#1|}
\newcommand{\ket}[1]{|#1\rangle}

\renewcommand{\ij}{{\langle \vec{i}, \vec{j} \rangle}}

\renewcommand{\H}{\hat{\mathcal{H}}}

\renewcommand{\c}{\hat{c}}

\newcommand{\cd}{\hat{c}^\dagger}

\newcommand{\bd}{\hat{b}^\dagger}

\renewcommand{\b}{\hat{b}}
\newcommand{\hd}{\hat{h}^\dagger}
\newcommand{\h}{\hat{h}}

\newcommand{\hc}{\text{h.c.}}

\usepackage{array}

\usepackage{cancel,ifthen}
\newcommand{\cmnt}[2][NoInPuT]{\ifthenelse{\equal{#1}{NoInPuT}}{}{{\color{red}\sout{#1}}} {\color{blue} #2}}

\usepackage{bm}	% bold in math mode
\renewcommand{\vec}[1]{\bm{#1}}

\LTcapwidth=\textwidth

\begin{document}
\normalem	% changes \emph back to normal after introducing ulem package.

\title{$\mathbb{Z}_2$ parton phases in the mixed-dimensional $t-J_z$ model}

\author{Fabian Grusdt}
\affiliation{Department of Physics and Arnold Sommerfeld Center for Theoretical Physics (ASC), Ludwig-Maximilians-Universit\"at M\"unchen, Theresienstr. 37, M\"unchen D-80333, Germany}
\affiliation{Munich Center for Quantum Science and Technology (MCQST), Schellingstr. 4, D-80799 M\"unchen, Germany}

\author{Lode Pollet}
\affiliation{Department of Physics and Arnold Sommerfeld Center for Theoretical Physics (ASC), Ludwig-Maximilians-Universit\"at M\"unchen, Theresienstr. 37, M\"unchen D-80333, Germany}
\affiliation{Munich Center for Quantum Science and Technology (MCQST), Schellingstr. 4, D-80799 M\"unchen, Germany}
\address{Wilczek Quantum Center, School of Physics and Astronomy, Shanghai Jiao Tong University, Shanghai 200240, China}


\date{\today}

\begin{abstract}

We study the interplay of spin- and charge degrees of freedom in a doped Ising antiferromagnet, where the motion of charges is restricted to one dimension. The phase diagram of this mixed-dimensional $ t - J_z$ model can be understood in terms of spin-less chargons coupled to a $\mathbb{Z}_2$ lattice gauge field. The antiferromagnetic couplings give rise to interactions between $\mathbb{Z}_2$ electric field lines which, in turn, lead to a robust stripe phase at low temperatures. At higher temperatures, a confined meson-gas phase is found for low doping whereas at higher doping values, a robust deconfined chargon-gas phase is seen which features hidden antiferromagnetic order. We confirm these phases in quantum Monte Carlo simulations. Our model can be implemented and its phases detected with existing technology in ultracold atom experiments. The critical temperature for stripe formation with a sufficiently high hole concentration is around the spin-exchange energy $J_z$, {\it i.e.}, well within reach of current experiments.

\end{abstract}

\maketitle

%%%%%%%%%%%%%%%%%%%%%%%%%%%%%%%%%%%%%%
\emph{Introduction.--}
%%%%%%%%%%%%%%%%%%
Ultracold atoms in optical lattices provide an excellent platform to perform analogue quantum simulations: they can mimic the behavior of tunable model Hamiltonians that are difficult or impossible to solve with current numerics. Since the advent of quantum simulators, an application to the 2D Fermi-Hubbard model has been a central goal: This model is believed to describe some of the most essential but theoretically poorly understood properties of strongly correlated electrons in the context of high-temperature superconductors. In the past years, significant steps have been taken towards simulating the Hubbard model, including the observation of long-range \cite{Mazurenko2017} and canted \cite{Brown2017} antiferromagnetism (AFM), bad metallic- \cite{Brown2019a} and spin- \cite{Nichols2018} transport, magnetic polarons \cite{Koepsell2019,Ji2020}, string patterns \cite{Chiu2019Science,Bohrdt2019NatPhys}, and in 1D spin-charge separation \cite{Hilker2017,Vijayan2020} and incommensurate magnetism \cite{Salomon2019}. Nevertheless, the critical temperatures of the expected ordered phase (stripes \cite{Zaanen2001}, superconductivity \cite{Qin2019}) are too low and have not yet been reached in ultracold fermion experiments.

%%%%%%%%%%%%%%%%%%%%%%%%%%%%%%%%%%%%%%%%%%%%%%%%%%%%%
\begin{figure}[t!]
\centering
\epsfig{file=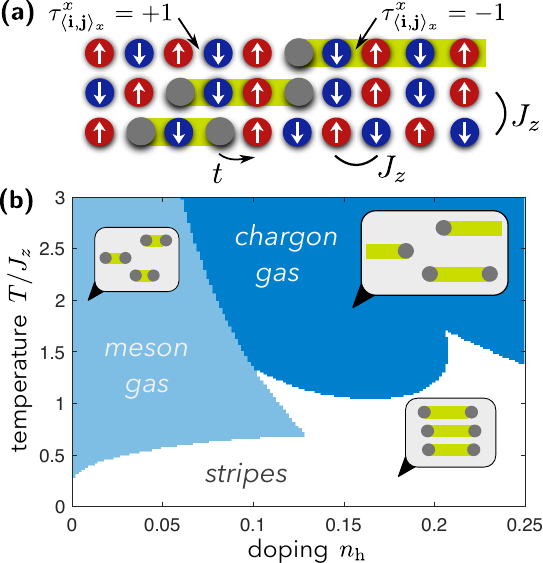, width=0.38\textwidth} $\quad$
\caption{The mix-D $t-J_z$ model with tunneling $t$ along $x$ and Ising couplings $J_z$ in both directions can be mapped to coupled 1D \Zt LGTs. (a) With a classical N\'eel background the \Zt electric field lines $\tau^x_{\ij_x}=-1$ denote regions where spins switch sub-lattice. (b) The phase diagram (here parton mean-field results for $t/J_z=3$ are shown) contains stripes, a confined meson gas, and a deconfined chargon gas. }
\label{figOverview}
\end{figure}
%%%%%%%%%%%%%%%%%%%%%%%%%%%%%%%%%%%%%%%%%%%%%%%%%%%%%

In this Letter we make use of the versatility of ultracold atoms to study a closely related cousin of the 2D Hubbard model. Its two main advantages  are: (i) significantly enhanced critical temperatures for the formation of stripe order amenable to quantum simulation; (ii)  thorough theoretical understanding and numerical control of the underlying physics. Both (i) and (ii) provide a promising starting point, in experiment and theory, for a systematic exploration of the 2D Hubbard model.

Specifically, we consider a $t-J_z$ model with mixed dimensionality \cite{Grusdt2018SciPost} as elucidated in Fig.~\ref{figOverview} (a),
\begin{equation}
\H = - t  \sum_{\sigma, \ij_x} \hat{\mathcal{P}}  \bigl( \cd_{\vec{i},\sigma} \c_{\vec{j},\sigma} + \hc \bigr) \hat{\mathcal{P}} + J_z  \sum_{\ij}  \hat{S}^z_{\vec{i}} \hat{S}^z_{\vec{j}}.
\label{eqModelDef}
\end{equation}
The dopants (holes) are free to move only along the $x$-direction, with tunneling rate $t$, while nearest-neighbor (NN) AFM Ising interactions between the spins, of strength $J_z $, are present along all dimensions of the lattice. In Eq.~\eqref{eqModelDef} $\ij$ denotes a pair of NN sites in a two-dimensional square lattice (every bond is counted once in the sum). Similarly, $\ij_x$ denotes a nearest neighbor bond oriented along the $x$-axis. We consider spin-$1/2$ particles $\c_{\vec{j},\sigma}$ with a hard-core constraint imposed by the projector $\hat{\mathcal{P}}$ onto the subspace without double occupancies. The statistics of the particles $\c_{\vec{j},\sigma}$ plays no role: By introducing Jordan-Wigner strings along the chains in $x$-direction one can switch between fermions and bosons.

%%%%%%%%%%%%%%%%%%%%%%%%%%%%%%%%%%%%%%
\emph{Symmetries and mapping to $\mathbb{Z}_2$ lattice gauge theory.--}
%%%%%%%%%%%%%%%%%%
Since holes cannot tunnel along $y$, their number $N^h_y$ is conserved in each chain $y$. In the following we restrict ourselves to equal doping $n_h$ in every chain, $N^h_y = n_{\rm h} L_x$ with the system size $L_{x(y)}$ along $x$ ($y$). 
In addition to the global $U(1)^{\otimes L_y}$ charge-conservation symmetries, and the conservation of total spin $\sum_{\vec{j}} S^z_{\vec{j}}$, the system exhibits \emph{hidden symmetries}. Namely, when the holes move they only change the positions of the spins in the 2D lattice, while it is impossible to permute their configurations within any given chain. This is formalized by the notion of squeezed space, introduced to describe 1D doped spin chains \cite{Ogata1990,Kruis2004a}: To this end Fock states $\otimes_y \ket{\sigma_{(1,y)}, ... ,\sigma_{(L_x-1,y)}, \sigma_{(L_x,y)} }$, with $\sigma_{\vec{j}} = \uparrow, h, \downarrow$ denoting local spin- and charge configurations, are re-labeled by $\otimes_y \ket{ \tilde{\sigma}_{(1,y)}, ... , \tilde{\sigma}_{(\tilde{L}_x,y)} } \otimes \hd_{(x_1,y)} ... \hd_{(x_{N^h_y},y)} \ket{0}$; Now $\tilde{\sigma}_{\vec{j}} = \uparrow, \downarrow$ denotes spins only on sites $\tilde{x}=1 ~...~ \tilde{L}_x = L_x-N^h_y$ and $\hd_{\vec{j}}$ creates a hard-core chargon with the same statistics as $\c_{\vec{j},\sigma}$ on the sites occupied by holes. The spin states in squeezed space are related to spins in the lattice by
\begin{equation}
 \tilde{\sigma}(\tilde{x},y) = \sigma \biggl( \tilde{x}+\sum_{j<\tilde{x}} n^h_{(j,y)} , y \biggr) \neq h,
 \label{eqDefSqzSpc}
\end{equation}
where $n^h_{\vec{j}}$ denotes the chargon occupation numbers. 

After this re-labeling, the eigenfunctions of \eqref{eqModelDef} become $\ket{\Psi} = \ket{\tilde{\Psi}} \otimes \ket{\Psi_c}$, where $\ket{\tilde{\Psi}} = \ket{ \{ \tilde{\sigma}_{\tilde{\vec{j}}} \}_{\tilde{\vec{j}}} }$ denotes a Fock configuration of spins in squeezed space and $\ket{\Psi_c}$ is a (generally correlated) chargon wavefunction \cite{Ogata1990}. Since we consider classical Ising interactions, every Fock configuration $\ket{\tilde{\Psi}}$ defines a separate hidden-symmetry sector of $\H$. In the following we restrict ourselves to N\'eel states in squeezed space: $\ket{\tilde{\Psi}} = \ket{\rm N} \equiv \ket{... \uparrow \downarrow \uparrow ...}$, with long-range antiferromagnetic correlations along $x$ and $y$ directions. 

If projected to the subspace $\ket{\tilde{\Psi}} = \ket{\rm N}$, the Hamiltonian for the chargons (with density $\hat{n}^h_{\vec{j}} = \hd_{\vec{j}} \h_{\vec{j}}$) becomes
\begin{equation}
\H = - t \sum_{\ij_x} \l \hd_{\vec{i}} \h_{\vec{j}} + \hc \r + \H_{\rm int}[ \left\{ \hat{n}^h_{\vec{j}} \right\} ],
\label{eqHeff}
\end{equation}
where the sign of the tunnelling term is irrelevant.

To express the non-local (but instantaneous) interaction energy $\H_{\rm int}[\{ \hat{n}^h_{\vec{j}} \} ]$ in a compact form, we introduce the following string operators,
\begin{equation}
\hat{\tau}^x_{\langle \vec{j}, \vec{j} + \vec{e}_x \rangle_x} = \prod_{\vec{i}: i_x \leq j_x} (-1)^{ \hat{n}^h_{\vec{i}}}.
\end{equation}
By definition, each pair of holes is connected by a string of link variables $\tau^x_\ij = -1$ [see Fig.~\ref{figOverview} (a)] and the following \Zt Gauss law is satisfied for all sites $\vec{j}$:
\begin{equation}
\hat{G}_{\vec{j}} \ket{\Psi} = \ket{\Psi}, \qquad \hat{G}_{\vec{j}} = \prod_{\vec{i}: \ij_x} \hat{\tau}^x_{\ij_x}.
\label{eqDefZ2Gauss}
\end{equation}
Owing to this Gauss law, the two link variables including a site $\vec{j}$ occupied by a spin $\sigma_{\vec{j}}$ are equal, $\tau^x_{\langle \vec{j}-\vec{e}_x, \vec{j} \rangle_x} = \tau^x_{\langle \vec{j}, \vec{j} + \vec{e}_x \rangle_x} = (-1)^{\pi_{\vec{j}}}$. Their value is given by the sub-lattice parity $\pi_{\vec{j}} = 0,1$ of this spin, {\it i.e.}, the number of times $\rm mod ~2$ the spin has switched sub-lattice (starting from a N\'eel state with all holes located on the right edge). 

The Ising interaction between neighboring spins $\ij_y$ along $y$ can be expressed in terms of the sub-lattice parities, $J_z \hat{S}^z_{\vec{i}} \hat{S}^z_{\vec{j}} = - J_z (-1)^{\pi_{\vec{i}} + \pi_{\vec{j}}} / 4$, since we use $\ket{\tilde{\Psi}} = \ket{\rm N}$. Along the chains each bond $\ij_x$ gives $J_z \hat{S}^z_{\vec{i}} \hat{S}^z_{\vec{j}} = - J_z/4$ unless one of the sites is occupied by a chargon.

We proceed by promoting the link variables to a \Zt lattice gauge theory (LGT)  subject to the \Zt Gauss law \eqref{eqDefZ2Gauss}. This requires adding a term $\hat{\tau}^z_{\ij_x}$ in the tunneling term in Eq.~\eqref{eqHeff} which correctly flips the sign of $\tau^x_{\ij_x}$, {\it i.e.}, $\hat{\tau}^z_{\ij_x} \ket{\tau^x_{\ij_x}} = \ket{- \tau^x_{\ij_x}}$. Note that the \Zt electric field $\hat{\tau}^x_{\ij_x}$ has a concrete physical meaning as it can be measured from the local spin configuration.  

Finally we arrive at the exact representation of Eq.~\eqref{eqModelDef}, in the sector $\ket{\tilde{\Psi}} = \ket{\rm N}$, by a \Zt LGT,
\begin{multline}
\H = - A \frac{J_z}{4}- t \sum_{\ij_x} \l \hd_{\vec{i}}  ~\hat{\tau}^z_{\ij_x} \h_{\vec{j}} + \hc \r + \frac{J_z}{2} \sum_{\vec{j}} \hat{n}^h_{\vec{j}} \\
- \frac{J_z}{4} \sum_{\ij_x} \hat{n}^h_{\vec{i}} \hat{n}^h_{\vec{j}} - \alpha \frac{J_z}{8} \sum_{\ij_y} (1-\hat{n}^h_{\vec{i}})  (1-\hat{n}^h_{\vec{j}}) \times \\
\times \biggl[ \hat{\tau}^x_{\langle \vec{i} - \vec{e}_x , \vec{i} \rangle_x} \hat{\tau}^x_{\langle \vec{j} - \vec{e}_x , \vec{j} \rangle_x}  +   \hat{\tau}^x_{\langle \vec{i} , \vec{i} + \vec{e}_x \rangle_x} \hat{\tau}^x_{\langle \vec{j}, \vec{j}+ \vec{e}_x \rangle_x} \biggr],
\label{eqZ2neelDef}
\end{multline}
where $A = L_x L_y$ is the total area. We introduced the dimensionless inter-chain coupling parameter $\alpha$, which is $\alpha=1$ for our model in Eq.~\eqref{eqModelDef}.

%%%%%%%%%%%%%%%%%%%%%%%%%%%%%%%%%%%%%%
\emph{Many-body phase diagram.--}
%%%%%%%%%%%%%%%%%%
Fig.~\ref{figOverview} (b) shows the phase diagram of the model in Eq.~\eqref{eqZ2neelDef} as a function of temperature $k_{\rm B}T$ and doping $n_h$. 
The phase boundaries are estimated using a parton-based mean-field description; note that our calculations for the stripe and meson regimes are restricted to low enough dopings to assume point-like constituents, which leads to unphysical cusps and re-entrances associated with stripes. See supplements for details. 
Each phase is also found in our quantum Monte Carlo (QMC) simulations.

%%%%%%%%%%%%%%%%%%%%%%%%%%%%%%%%%%%%%%%%%%%%%%%%%%%%%
\begin{figure}[t!]
\centering
\epsfig{file=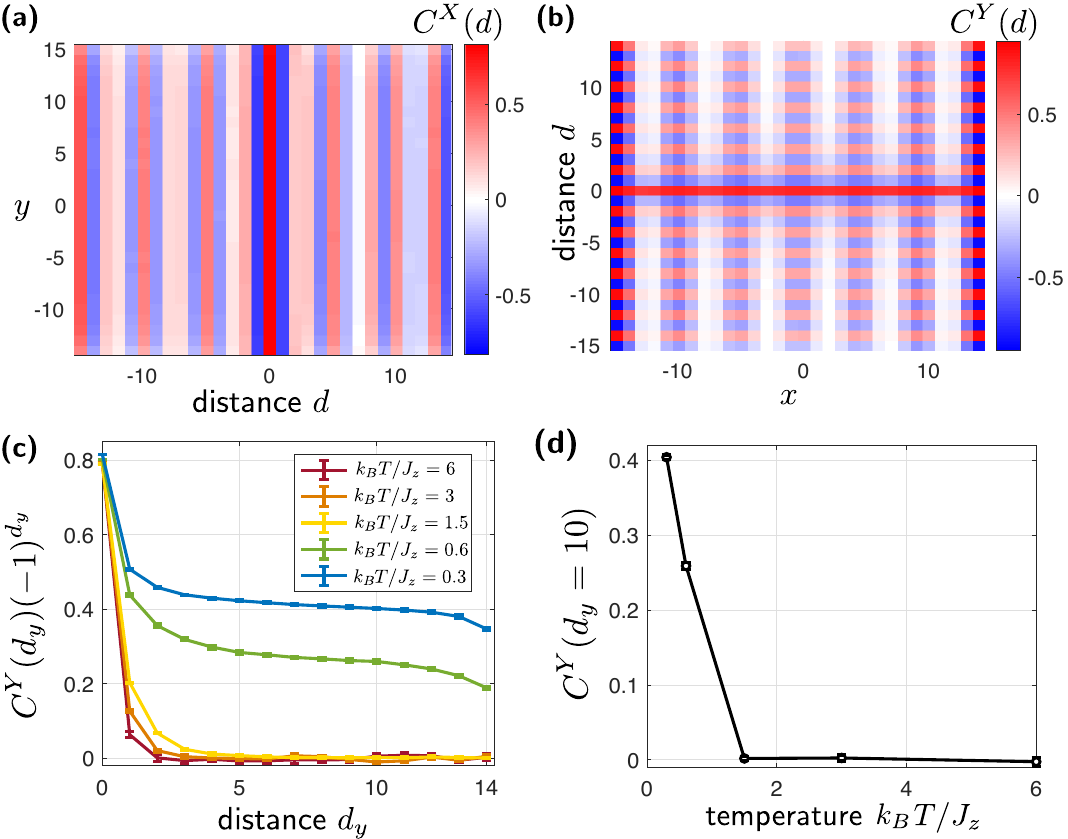, width=0.499\textwidth}
\caption{Stripe formation: QMC simulations of Eq.~\eqref{eqZ2neelDef} reveal the onset of stripe order at low temperatures. We show $C^X(d)$ in (a) [$C^Y(d)$ in (b)] relative to the central column [chain] at $d=0$ for $k_B T = 0.6 J_z$. (c) For different temperatures we show how long-range AFM spin-correlations $(-1)^d C^Y(d)$ develop perpendicular to the chains; $C^Y(d)$ is measured relative to the central chain. The correlator at a large distance $d=10$ is shown in (d). We consider a $30 \times 30$ system (open boundaries), $6$ holes per chain, and $t/J_z=3$.} 
\label{figStripes}
\end{figure}
%%%%%%%%%%%%%%%%%%%%%%%%%%%%%%%%%%%%%%%%%%%%%%%%%%%%%

For the ground state ($T=0$) we predict a vertical stripe phase, characterized by charge modulations with wavelength $\lambda = 1/n_{\rm h}$. The \Zt electric field changes sign across each stripe, respecting the \Zt Gauss law. 

As a result, incommensurate long-range spin correlations are found along $x$, see Fig.~\ref{figStripes} (a):
\begin{equation}
C^X(d) \equiv 4 \left\langle \hat{S}^z_{\vec{j}} \hat{S}^z_{\vec{j}+d \vec{e}_x} \right\rangle \simeq \nu^X_{\rm S} \cos \bigl( \pi (1+ n_{\rm h}) d \bigr),~ d \to \infty.
\label{eqDefCX}
\end{equation}
The binding mechanism into stripes can be readily understood from Eq.~\eqref{eqZ2neelDef}: The interactions of the \Zt electric field lines favor alignment of the latter along $y$, which is achieved by creating strong charge correlations along the $y$-direction. Such localization along $y$ is cheap due to the absence of chargon tunneling in this direction. On the other hand, strong anti-bunching along $x$ allows each chargon to delocalize as much as possible, in direct competition with the attraction of \Zt electric field lines.

As shown in Fig.~\ref{figStripes} (b), stripes are indeed characterized by long-range AFM order in the $y$-direction (corresponding to aligned \Zt electric field lines):
\begin{equation}
C^Y(d) \equiv 4 \left\langle \hat{S}^z_{\vec{j}} \hat{S}^z_{\vec{j}+d \vec{e}_y} \right\rangle \simeq \nu^Y_{\rm S} (-1)^d,~ d \to \infty.
\label{eqDefCY}
\end{equation}
Numerically we find that long-ranged correlations $C^Y(d)$ develop below a non-zero critical temperature $T_{\rm S}>0$. Our QMC simulations in Fig.~\ref{figStripes} (c) and (d) show that $k_B T_{\rm S} \approx 1.0(5) J$ for the chosen value of $t/J_z=3$ and $20\%$ hole doping for linear system size $L=15$.

Within each chain our system has a conserved number of holes, associated with separate $U(1)$ symmetries. In the long-wavelength limit, the corresponding effective field theory describes a $U(1)$ symmetric field without quantum fluctuations of the charge along $y$. Integrating out thermal fluctuations at temperatures $k_B T>0$ yields an effective action of a $1+1$ dimensional quantum system. With the global $U(1)$ symmetry along $y$, we thus expect power-law correlations along $x$ and $y$ in the stripe phase: Below the critical temperature for stripe formation, $T_{\rm S}>0$, these replace the infinite-range correlations Eqs.~\eqref{eqDefCX}, \eqref{eqDefCY} expected in the true ground state. 

We find that our finite-size simulations with open boundaries are consistent with very weak power-law correlations $C^Y(d)$ when $0 < T \lesssim T_{\rm S}$. The detailed nature of the transition at $T_{\rm S}$ remains a subject of future investigation, but we expect it to be in BKT class. 

At higher temperatures and beyond a rather small critical doping value $n_{\rm h} \geq n_{\rm h}^{c}(T)$ we predict a \emph{chargon gas}. It has no long-range AFM order in either direction, $C^X(d), C^Y(d) \to 0$ as $d \to \infty$. The loss of antiferromagnetism is entirely due to chargon dynamics, however: in squeezed space the spin wavefunction is still the classical N\'eel state. Hence the chargon gas is characterized by its \emph{hidden AFM order}, which manifests itself in the non-local string correlations defined by the \Zt Gauss law \eqref{eqDefZ2Gauss}. Related string correlations have been observed in 1D Hubbard models \cite{Endres2011,Hilker2017} and are commonly used to characterize topological order in 1D systems \cite{denNijs1989,Fazzini2019}.

In contrast to the stripe phase, the chargon gas is characterized by a \emph{disordered} \Zt electric field:
\begin{equation}
 e_{\ij_x} \equiv \langle \hat{\tau}^x_{\ij_x} \rangle = 0.
\end{equation}
Chargons are hence deconfined and form a gapless phase \cite{Borla2019}, corresponding to free fermionic holes at the mean-field level.

Finally, at very low doping $n_{\rm h} < n_{\rm h}^{c}(T)$ but above the critical temperature $T > T_{\rm S}(n_{\rm h})$ for stripe formation, we predict a \emph{meson gas}. It is characterized by a uniform \Zt electric order parameter
\begin{equation}
e_{\ij_x} \equiv \langle \hat{\tau}^x_{\ij_x} \rangle = \nu_{\rm cc} \neq 0.
\end{equation}
This should be contrasted to the $T=0$ stripe phase with incommensurate magnetism, where $e_{\ij_x} \neq 0$ is modulated in space with a wavelength $\lambda = 2 / n_{\rm h}$, such that $\sum_{\ij}  e_{\ij_x} =0$; in the stripe phase at $0 < T < T_{\rm S}$ the thermal average $e_{\ij_x} =0$ is expected to be strictly zero in the thermodynamic limit. As a direct consequence of $\nu_{\rm cc} \neq 0$, the meson gas has commensurate long-range AFM order along both directions,
\begin{equation}
 C^X(d), C^Y(d) ~ \to~  (-1)^d \nu_{\rm cc}^2 , \qquad d \to \infty.
\end{equation}

%%%%%%%%%%%%%%%%%%%%%%%%%%%%%%%%%%%%%%%%%%%%%%%%%%%%%
\begin{figure*}[t!]
\centering
\epsfig{file=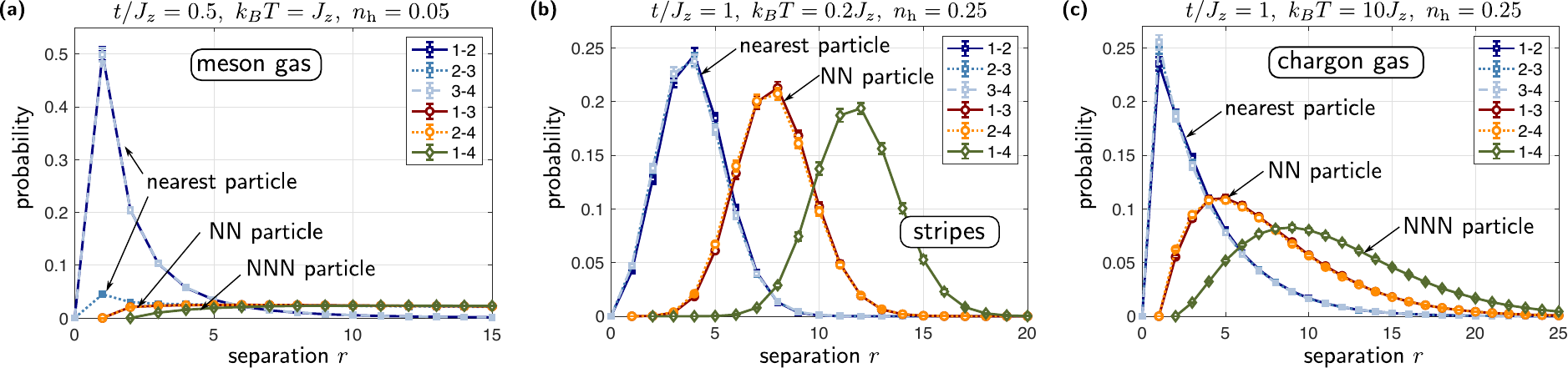, width=0.99\textwidth}
\caption{Chargon distance histograms. We plot the distributions $p_{n,m}(r)$ of separations $r$ between chargons number $n$ and $m$ in the chains, counting from the left. In the meson gas phase (a) $p_{1,2}(r) = p_{3,4}(r)$ is significantly broader than $p_{2,3}(r)$, a direct indication for pairing. In the stripe phase (b), $p_{1,2}(r) = p_{2,3}(r) = p_{3,4}(r)=...$ are equal and all distributions are narrow, indicating localization of chargons into stripes. (c) In the chargon gas phase, $p_{1,2}(r) = p_{2,3}(r) = p_{3,4}(r)=...$ and all distributions feature long tails. In all simulations we used an $80 \times 10$ system (other parameters as indicated).} 
\label{figHistograms}
\end{figure*}
%%%%%%%%%%%%%%%%%%%%%%%%%%%%%%%%%%%%%%%%%%%%%%%%%%%%%

Physically, the meson gas can be understood as a paired phase of chargons. The \Zt electric string connecting two chargons is associated with a linear string tension $\propto \nu_{\rm cc}$, which precludes one-chargon excitations in the thermodynamic limit; i.e. the meson gas corresponds to a confined phase which has, even in the zero-temperature limit, $\langle \hd_{\vec{j}} \bigl( \prod_{\vec{j} \leq \langle \vec{i}, \vec{k} \rangle_x \leq \vec{j}+d \vec{e}_x} \hat{\tau}^z_{\langle \vec{i}, \vec{k} \rangle_x} \bigr) \h_{\vec{j} + d \vec{e}_x} \rangle \simeq e^{- \eta d}$ for $d \to \infty$. Due to the restriction of chargon dynamics along one direction, the meson gas also corresponds to a Luttinger liquid with fractionalized excitations in the zero-temperature limit \cite{Borla2019}. 

To identify the meson gas phase in our QMC numerics, we calculate histograms of chargon separations in Fig.~\ref{figHistograms}. The hallmark of chargon-chargon meson formation is a narrow distribution $p_{2n-1,2n}(r)$ of separations $r$ between chargon numbers $2n-1$ and $2n$, with $n=1,2,...$ and counting from the left edge, and a broader and different distribution $p_{2n,2n+1}(r)$ between chargons $2n$ and $2n+1$. This feature is clearly visible in Fig.~\ref{figHistograms} (a) in the expected low-doping regime, where we also find a non-vanishing \Zt electric order parameter, 
% 0.88421197466 +- 0.000744905636729
$\langle \hat{\tau}^x_{\ij_x} \rangle = 0.8842(8)$. In the other two phases, the histograms show significantly different features, see Fig.~\ref{figHistograms} (b), (c). 

In the phase diagram, the meson gas is associated with an unusual re-entrant behavior as one increases temperature along a line of constant, but small, doping: at $T=0$ the system has incommensurate long-range AFM correlations, which we predict to be destroyed by thermal fluctuations of the stripes when $0 < T < T_{\rm S}$. When the meson gas phase is entered for $T > T_{\rm S}$, true long-range AFM correlations are restored. This counter-intuitive behavior is possible since AFM correlations are merely hidden in the intermediate fluctuating stripe regime. 

Finally, we want to make a connection with Ref.~\cite{Grusdt2018SciPost}, where a single mobile dopant but with $SU(2)$ invariant Heisenberg interactions has been studied. It was found that  the hole forms a magnetic polaron \cite{Kane1989,Sachdev1989,Martinez1991,Koepsell2019} which can be understood as a meson-like bound state of a spinon and a chargon \cite{Beran1996} connected by a geometric string of displaced spins \cite{Grusdt2018SciPost}. Our meson phase is an analog of this finding but at finite hole concentration and for Ising-type interactions.

%%%%%%%%%%%%%%%%%%%%%%%%%%%%%%%%%%%%%%%%
\emph{Methods.-- }
Our calculations are based on a number of different but standard techniques such as bosonization, mean-field parton theory, and QMC simulations.

In order to work  with a 1D field theory amenable to bosonization, our crucial approximation is the decoupling ansatz
\begin{equation}
 \hat{\tau}^x_{\langle \vec{i} , \vec{i} + \vec{e}_x \rangle_x} \hat{\tau}^x_{\langle \vec{j}, \vec{j}+ \vec{e}_x \rangle_x} \approx  V_{\rm MF}(i_x) [ \hat{\tau}^x_{\langle \vec{i} , \vec{i} + \vec{e}_x \rangle_x} + \hat{\tau}^x_{\langle \vec{j}, \vec{j}+ \vec{e}_x \rangle_x} ],
\end{equation}
for $\ij_y$ NN along $y$, i.e. $i_x = j_x$. The different phases correspond to different solutions for $V_{\rm MF}(i_x)$. These approximations are justified because we find the same phases in the quantum Monte Carlo simulations. We find the critical Luttinger parameter below which the ground state forms stripes to be large, $K_c=8$. We refer to the supplementary for details.

%%%%%%%%%%%%%%%%%%%%%%%%%%%%%%%%%%%%%%
\emph{Discussion and outlook.--}
In summary, we showed that the mix-D $t-J_z$ model can be directly mapped onto a \Zt LGT. The many-body phase diagram of our model features in the ground state a stripe phase where the holes form vertical walls. Above a critical temperature  $T_{\rm S}$ , but  within the N\'eel  \Zt gauge sector (which has the lowest energy at zero doping), we find two gaseous phases: a confined meson gas, with long-range AFM order, and a  deconfined chargon gas with hidden AFM correlations.

Experimentally, the model Eq.~\eqref{eqModelDef} can be realized in the large $U/t$-limit of a bosonic Hubbard model with a strong tilt $\Delta \gg t$ along the $y$-direction: The strong tilt suppresses resonant tunnelling of dopants along $y$, whereas the super-exchange mechanism remains intact in both directions \cite{Grusdt2018SciPost,Dimitrova2019}; to obtain AFM Ising interactions, one can use spin-dependent scattering lengths \cite{Duan2003,Dimitrova2019}. Rydberg atoms, which have already demonstrated Ising spin systems \cite{Schauss2012,Schauss2015,Labuhn2016,Bernien2017,GuardadoSanchez2017,Lienhard2017}, are an alternative option: By using multiple hyperfine levels to encode both spin and charge degrees of freedom, our mix-D $t-J_z$ model should also be realizable; see also Ref.~\cite{Gorshkov2011tJ} for a discussion of generic $t-XYZ$ models in polar molecules, and Refs.~\cite{Zohar2017,Schweizer2019a,Barbiero2019,Halimeh2019} for direct implementations of \Zt LGTs. For all systems, we propose to start from a classical N\'eel state without holes, which can be doped with mobile charges e.g. by adiabatic deformations of the trapping potentials. This should guarantee that thermal fluctuations outside the gauge sector of our \Zt LGT are negligible.

In spite of the overwhelming simplifications of our model,  the presence of a stripe phase and a confinement-to-deconfinement transition at elevated temperatures draws one's attention to the cuprates. A goal for future investigations is to study related models which are more closely related to the 2D $t-J$ model: as a first step, other gauge sectors with domain walls in squeezed space -- corresponding to spinons -- can be considered. By replacing Ising interactions with $SU(2)$ invariant Heisenberg couplings, a much richer model is expected and it remains to be seen if any connections to  \Zt LGTs can be drawn. Finally, the goal is to include charge dynamics along the second direction: this may provide an adiabatic route to the stripe phase observed in cuprates.

%%%%%%%%%%%%%%%%%%%%%%%%%%%%%%%%%%%%%%
\emph{Acknowledgements.--}
We thank J. Amato-Grill, L. Barbiero,  I. Bloch, A. Bohrdt, U. Borla, M. Buser, P. Cubela, E. Demler, I. Dimitrova, S. Eggert, N. Goldman, M. Greiner,  C. Gross, N. Jepsen, M. Kebric, J. Koepsell,  S. Moroz, M. Punk, G. Salomon, U. Schollw\"ock, T. Shi, R. Verresen, Z. Zhu for fruitful discussions. 
Numerical data for this paper is available under \url{https://github.com/LodePollet/QSIMCORR}.
We acknowledge funding by the Deutsche Forschungsgemeinschaft (DFG, German Research Foundation) under Germany's Excellence Strategy -- EXC-2111 -- 390814868 and via Research Unit FOR 2414 under project number 277974659. LP is supported by FP7/ERC Consolidator Grant  No. 771891 (QSIMCORR).

%%%%%%%%%%%%%%%%%%%%%%%%%%%%%%%%%%%%%%%%%%%%%%%%%%%%%
\begin{figure*}[t!]
\centering
\epsfig{file=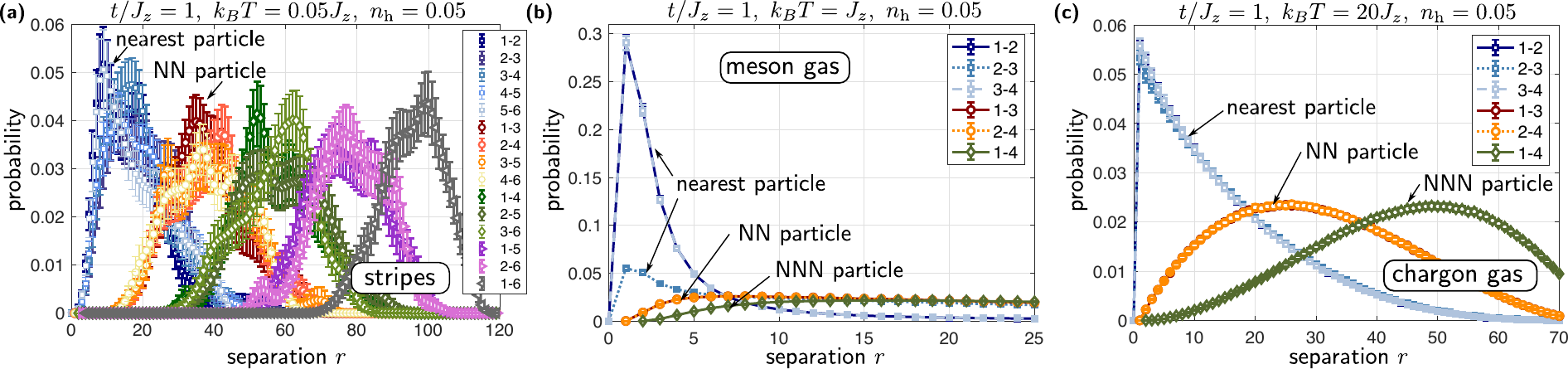, width=0.99\textwidth}
\caption{Chargon distance histograms. We plot the distributions $p_{n,m}(r)$ of separations $r$ between chargons number $n$ and $m$ in the chains, counting from the left. For all plots, $t/J_z=1$ and we considered a doping of $n_{\rm h} = 5 \%$ [6 holes per chain in a $120 \times 60$ periodic system in (a) and 4 holes per chain in a $80 \times 10$ open system in (b) and (c)]. From (a) to (c) the temperature $k_{B}T$ is increased, showing stripes at low $T$ (a), the meson gas at intermediate $T$ (b) and the chargon gas at high $T$ in (c).} 
\label{figHistogramsSM}
\end{figure*}
%%%%%%%%%%%%%%%%%%%%%%%%%%%%%%%%%%%%%%%%%%%%%%%%%%%%%

%\newpage~
\newpage

%%%%%%%%%%%%%%%%%%%%%%%%%%%%%%%%%%%%%%%%%%%%%%%%%%%%%
\section{Supplementary}
%%%%%%%%%%%%%%%%%%%%%%%%%%%%%%%%%%%%%%%%%%%%%%%%%%%%%

%%%%%%%%%%%%%%%%%%%%%%%%%%%%%%%%%%%%%%
\subsection{Bosonization}
%%%%%%%%%%%%%%%%%%
To understand the many-body phase diagram, we bosonize the \Zt LGT chains along $x$. As a starting point we consider decoupled chains ($\alpha=0$ in Eq.~\eqref{eqZ2neelDef}) and study when interactions between the chains ($\alpha \neq 0$) become relevant. For $\alpha=0$ the chargons form a Luttinger liquid with $K \approx 1$ \cite{Prosko2017} which corresponds to the chargon gas phase; the weak NN attraction $-J_z \hat{n}_{\vec{i}} \hat{n}_{\vec{j}} / 4$ is expected to lead to slight deviations from a free Fermi gas with $K \gtrsim 1$ \cite{Giamarchi2003}. 

To continue working with a 1D field theory amenable to bosonization, we make a mean-field ansatz as written already in the main text, and write
\begin{equation}
 \hat{\tau}^x_{\langle \vec{i} , \vec{i} + \vec{e}_x \rangle_x} \hat{\tau}^x_{\langle \vec{j}, \vec{j}+ \vec{e}_x \rangle_x} \approx  V_{\rm MF}(i_x) [ \hat{\tau}^x_{\langle \vec{i} , \vec{i} + \vec{e}_x \rangle_x} + \hat{\tau}^x_{\langle \vec{j}, \vec{j}+ \vec{e}_x \rangle_x} ],
\end{equation}
for $\ij_y$ NN along $y$, i.e. $i_x = j_x$. The different phases we predict correspond to qualitatively different mean-field solutions $V_{\rm MF}(i_x)$.

We start with stripes at zero temperature, characterized by periodic charge modulations with wavenumber $k=2 \pi n_{\rm h}$. Since the \Zt electric field changes sign across every stripe, $V_{\rm MF}(x+1/n_{\rm h}) = - V_{\rm MF}(x)$. Hence we obtain the following Fourier expansion,
\begin{equation}
V_{\rm MF}(x) = \sum_{m=0}^{\infty} V_{2m+1} \cos \bigl( (2m+1) \pi n_{\rm h} x \bigr).
\label{eqVMFexp}
\end{equation}
Using the \Zt Gauss law \eqref{eqDefZ2Gauss} we can express $\hat{\tau}^x_{\langle \vec{j} , \vec{j} + \vec{e}_x \rangle_x} = \prod_{i_x \leq j_x} \cos( \pi \hat{n}^h_{(i_x,j_y)} )$ in terms of the chargon density. From the bosonization formula $\hat{n}^h_{\vec{j}} \simeq n_{\rm h} - \partial_x \hat{\phi}(x_{\vec{j}}) / \pi$ we find that a non-oscillating term in $x$ survives in the resulting effective action per chain,
\begin{equation}
\mathcal{S} = \mathcal{S}_0 - \frac{\alpha J_z}{4} (1-n_{\rm h})^2 V_1 \int dx d\tau~ \cos(\phi(x,\tau)),
\label{eqEffAction}
\end{equation}
where $\mathcal{S}_0$ is the free action with Luttinger parameter $K$.

The interaction in Eq.~\eqref{eqEffAction} is relevant for $K < 8$, showing that our model with $K \approx 1$ for $t \gtrsim J_z$ should form stripes. The comparatively large critical value of $K_c=8$ indicates that long-ranged attractive interactions would be required to de-stabilize the stripe phase. The effective action in Eq.~\eqref{eqEffAction} highlights another peculiarity of the \Zt LGT: the gapped elementary excitations correspond to $2\pi$ phase slips of the field $\hat{\phi}(x)$, which in turn correspond to \emph{pairs} of two chargons. This demonstrates that chargon excitations on top of the stripe ground state are confined and form mesons. 

The meson gas phase corresponds to a uniform mean-field solution $V_{\rm MF}(x) = \nu_{\rm cc} \neq 0$. In this case only $V_0 \neq 0$ in Eq.~\eqref{eqVMFexp} and the resulting effective action $\mathcal{S}_{\alpha}$ has a term oscillating with $\cos(\pi n_{\rm h} x)$, making the inter-chain coupling $\alpha$ irrelevant in this phase -- see Ref.~\cite{Borla2019}.

%%%%%%%%%%%%%%%%%%%%%%%%%%
\subsection{QMC simulations}
%%%%%%%%%%%%%%%%%%%%%%%%%%

The quantum Monte Carlo simulations are based on a sampling of the path integral representation of the partition function with the help of worm-type updates \cite{Prokofev1998qmc}, here in the variant of Ref~\cite{Pollet2007}. The structure of our Hamiltonian in Eq.~\eqref{eqModelDef} is similar to a two-component Bose-Hubbard system with nearest-neighbor density-density interactions, which is standard. The  N\'eel gauge sector is implemented via the initial configuration as well as making sure that the imaginary time difference between the worm ends never exceeds $\beta/2$ in absolute value. This also makes the simulations canonical.

We observed rather large autocorrelation and thermalization times in the simulations. We attribute this to the physics of the phases we found: In the stripe phase, effectively shifting the position of the hole wall has a large barrier, whereas in the meson phase the binding energy of two holes makes it likewise difficult to move such pairs around.

In Fig.~\ref{figHistogramsSM} we show more QMC results for the chargon distance histograms. We consider $t/J_z=1$ and fix the doping to $n_{\rm h} = 5 \%$. For the lowest temperature (a) we find no significant differences between the distributions $p_{1,2}(r)$ and $p_{2,3}(r)$ of separations $r$ between particles $1$ and $2$ (respectively, $2$ and $3$). For intermediate temperatures $k_{B} T = J_z$ (b) we find that particles $2$ and $3$ are significantly further apart than $1$ and $2$ or $3$ and $4$. This is a signature of the meson gas phase. At higher temperatures $k_B T = 20 J_z$ in (c) we find a significantly broader distribution where the distances between particles $1$ and $2$, $2$ and $3$, and $3$ and $4$ behave very similarly. These features indicate a chargon gas phase.

\begin{figure}[t!]
\centering
\epsfig{file=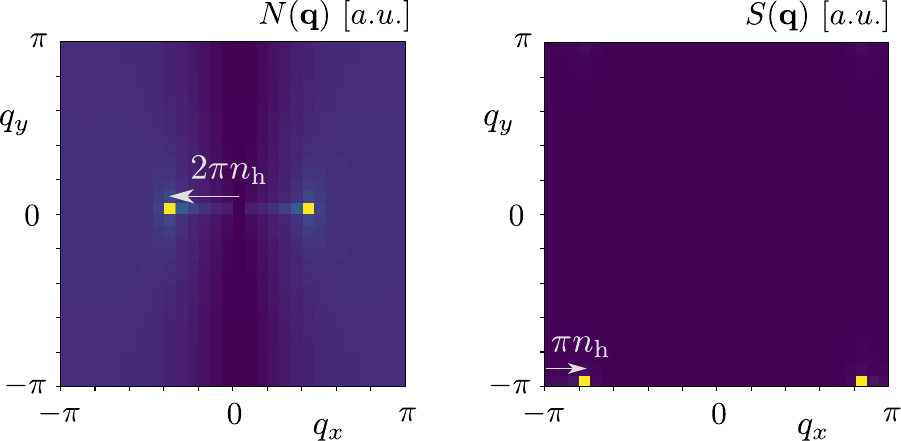, width=0.49\textwidth}
\caption{Static charge (left) and spin (right) structure factors $N(\vec{q})$ ($S(\vec{q})$, respectively) in the stripe phase. Parameters are the same as in Fig.~\ref{figStripes} in the main text: we consider a $30 \times 30$ system with open boundary conditions,  $t/J_z=3$,  at a doping of $20 \%$ (6 holes per chain) and temperature $k_B T = 0.3 J_z$.} 
\label{figStructureFactors}
\end{figure}

In Fig.~\ref{figStructureFactors} we show static spin and charge structure factors in the stripe phase. As expected we observe sharp delta-peaks at $\pm 2 \pi n_{\rm h}$ in the charge sector, and $\pm \pi n_{\rm h}$ in the spin sector relative to the $(0,0)$ and $(\pi, \pi)$ values expected in the hole-free background, respectively -- another hallmark of stripe formation. We also checked (not shown) that in the meson phase the static charge and spin structure factors show peaks at $(0,0)$ and $(\pi, \pi)$, respectively -- as expected.

%%%%%%%%%%%%%%%%%%%%%%%%%%%%%%%%%%%%%%%%%%%%%%%%%%%%%
%%%%%%%%%%%%%%%%%%%%%%%%%%%%%%%%%%%%%%%%%%%%%%%%%%%%%

%%%%%%%%%%%%%%%%%%%%%%%%%%
\subsection{Microscopic parton theory}
%%%%%%%%%%%%%%%%%%%%%%%%%%

In this section we introduce the microscopic mean-field descriptions of parton phases which we used to determine the phase diagram in Fig.~\ref{figOverview} (b) of the main text. We focus on the low doping regime $n_{\rm h} \ll 1$, except for the chargon gas phase. Our starting point is the Hamiltonian Eq.~\eqref{eqZ2neelDef} in the gauge sector $\hat{G}_{\vec{j}} \ket{\psi} = \ket{\psi}$ for all $\vec{j}$, and we ignore finite-size effects in the following. The details of our calculation in Fig.~\ref{figOverview} (b) of the main text are described in \ref{SecSMresults}.

%%%%%%%%%%%%%%%%
\subsubsection{Mean-field theory of stripes}
%%%%%%%%%%%%%%%%
Our strategy in this section is to describe how individual stripes along $y$ can form. These stripes contain exactly one hole per chain and stripe; they are expected to have long-range charge correlations along $y$. The basic binding mechanism is provided by the attractive interaction of parallel \Zt electric field lines. For simplicity we neglect interactions between stripes at different positions $x_1$ and $x_2$. To describe the individual stripes, we focus on the simplest case with exactly one hole per chain.

\emph{Zero temperature.--}
We describe the one-stripe ground state by a variational Gutzwiller wavefunction,
\begin{equation}
	\ket{\Psi_{\rm S}} = \prod_{y=1}^{L_y} \ket{\phi^{(0)}}_y.
	\label{eqDefMFstripes}
\end{equation}
It represents a product of identical chargon wavefunctions $\ket{\phi^{(0)}}_y$ per chain $y$. We can express the latter as
\begin{multline}
 	\ket{\phi}_y = \sum_{x=-L_x/2}^{L_x/2} \phi^{(0)}_x ~  \hd_{(x,y)} \prod_{\langle i,j \rangle \geq x} \hat{\tau}^z_{\langle (i,y) , (j,y) \rangle_x} \ket{0} \\
	\equiv \sum_{\ell} \phi^{(0)}_\ell ~ \ket{\ell}_y. \qquad \qquad \qquad \qquad \qquad \qquad
	\label{eqDefPhiStripe}
\end{multline}
where $\ket{0}$ denotes the undoped N\'eel state. Assuming that the stripe is centered around $x=0$, we denote the basis states by $\ket{\ell}_y$ in the second line of Eq.~\eqref{eqDefPhiStripe}: here $\ell$ can be understood as the length of the \Zt electric string measured relative to center of the stripe. 

\begin{figure}[t!]
\centering
\epsfig{file=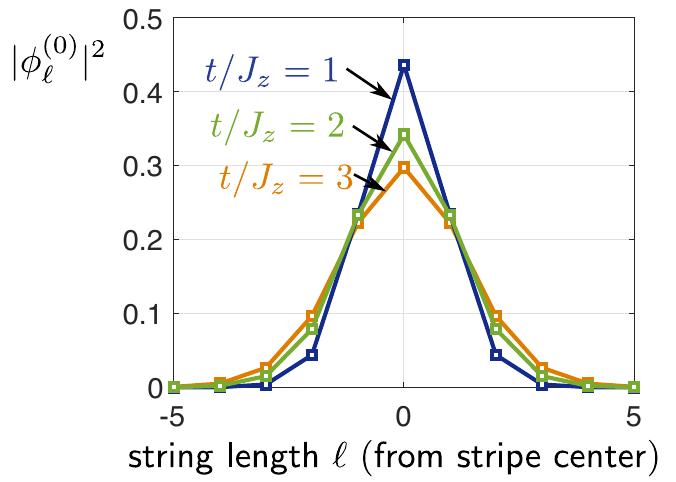, width=0.35\textwidth} $\qquad$
\caption{String length distribution $|\phi_\ell^{(0)}|^2$ in the stripe phase at zero temperature, measured from the stripe center. We used $\ell_{\rm max}=30$ in the calculation.} 
\label{figEllS}
\end{figure}

To approximate the ground state of the stripe, we optimize the wavefunction $\phi^{(0)}_\ell$ in order to minimize the variational energy:
\begin{multline}
\frac{ \langle \H \rangle_0 }{ L_y } = - L_x \frac{J_z}{2} - t \sum_\ell \l \phi^{(0)*}_{\ell+1}   ~ \phi^{(0)}_\ell + {\rm c.c.} \r +\\
 + \sum_{\ell, r} |\phi^{(0)}_\ell|^2 ~  |\phi^{(0)}_r|^2 ~ V_{\rm pot}(\ell-r),
 \label{eqHvarStripes}
\end{multline}
where the inter-chain potential is given by
\begin{equation}
 V_{\rm pot}(\ell) = \frac{J_z}{2} + \frac{J_z}{4} \delta_{\ell,0} + \frac{J_z}{2} |\ell|.
 \label{eqVellDef}
\end{equation}
The optimization problem is easy to solve numerically, by introducing a maximum string length $|\ell| \leq \ell_{\rm max}$ in the one-particle Hilbertspace $\ket{\ell}$. In Fig.~\ref{figEllS} we show the probability amplitude of the optimized wavefunction for different values of $t/J_z$.

From the variational ground state energy $ \langle \H \rangle_0$ we obtain the energy per hole in one stripe, which we measure relative to the undoped N\'eel state:
\begin{equation}
\epsilon_0^{\rm S} =  \frac{\langle \H \rangle_0}{L_y} + L_x \frac{J_z}{2}.
\label{eqE0SperHole}
\end{equation}

%%%%%%%%%%%%%%%%%%%%%%%%%%%%%%%%%%%%%%%%%%%%%%%%%%%%%
%%%%%%%%%%%%%%%%%%%%%%%%%%%%%%%%%%%%%%%%%%%%%%%%%%%%%

Instead of variationally optimizing $\phi_\ell^{(0)}$ to minimize the overall energy, we can derive a self-consistency mean-field equation: Averaging over neighboring chains, the effective mean-field Hamiltonian per chain becomes
\begin{equation}
 \H_{\rm MF} =  - L_x \frac{J_z}{2} + \sum_\ell \biggl[ -t \bigl( \ket{ \ell+1 } \bra{ \ell } + \hc \bigr) + \ket{\ell} \bra{\ell} V_{\rm eff}(\ell) \biggr],
 \label{eqDefHMFstripes}
\end{equation}
where the self-consistent mean-field potential is
\begin{equation}
 V_{\rm eff}(\ell) = 2 \sum_r |\phi_r^{(0)}|^2 V_{\rm pot}(r-\ell).
\end{equation}
We find that the variationally optimized solution is an eigenstate of the mean-field Hamiltonian.

\emph{Phonon spectrum.--}
To describe the effects of thermal fluctuations later on, the excitation spectrum of stripes is required. To this end we make a variational ansatz 
\begin{equation}
 \ket{\Psi_{\rm ph}(k_y)} =  \sum_y  \frac{e^{i k_y y}}{\sqrt{L_y}} \biggl[ \prod_{y' \neq y} \ket{\phi^{(0)}}_{y'} \biggr] \otimes \ket{\phi^{(1)}}_y
 \label{eqDefPhonon}
\end{equation}
with momentum $k_y$ along the $y$-direction. Here $\ket{\phi^{(0)}}_{y'}$ is the optimized ground state wavefunction obtained by minimizing $\langle \H \rangle$ in Eq.~\eqref{eqHvarStripes}. The resulting phonon dispersion is given by
\begin{equation}
 \omega_{\rm ph}(k_y) = \bra{\Psi_{\rm ph}(k_y)} \H \ket{\Psi_{\rm ph}(k_y)} - \bra{\Psi_{\rm S}} \H \ket{\Psi_{\rm S}}.
 \label{eqOmegaPhonon}
\end{equation}

Since components in $\ket{\phi^{(1)}}$ which are not orthogonal to $\ket{\phi^{(0)} }$ vanish in Eq.~\eqref{eqDefPhonon}, we may assume
\begin{equation}
 \bra{ \phi^{(1)} } \phi^{(0)} \rangle = 0.
\end{equation}
Using this relation, Eq.~\eqref{eqOmegaPhonon} yields
\begin{equation}
\omega_{\rm ph}(k_y) = \Delta_{\rm S}(k_y) + 2 J_{\rm ex}(k_y) ~ \cos(k_y),
\end{equation}
where (lower indices denote chain numbers):
\begin{flalign*}
 \Delta_{\rm S}(k_y) &= \epsilon^{(1)}_{\rm MF}(k_y) - \epsilon^{(0)}_{\rm MF}, \\
 \epsilon^{(1)}_{\rm MF}(k_y) & = 2 \bra{\phi_1^{(0)}} \bra{\phi_2^{(1)}}   ~ \hat{V}~  \ket{\phi_2^{(1)}} \ket{\phi_1^{(0)}} + \bra{\phi^{(1)}} \hat{T} \ket{\phi^{(1)}}, \\
 \epsilon^{(0)}_{\rm MF}(k_y) & = 2 \bra{\phi_1^{(0)}} \bra{\phi_2^{(0)}}   ~ \hat{V}~  \ket{\phi_2^{(0)}} \ket{\phi_1^{(0)}} + \bra{\phi^{(0)}} \hat{T} \ket{\phi^{(0)}}, \\
 J_{\rm ex} &= \bra{\phi_1^{(0)}} \bra{\phi_2^{(1)}}   ~ \hat{V}~  \ket{\phi_2^{(0)}} \ket{\phi_1^{(1)}}, 
\end{flalign*}
and we defined
\begin{flalign*}
 \hat{V} &= \sum_{\ell_1,\ell_2} \ket{\ell_1,\ell_2} \bra{\ell_1, \ell_2} ~ V_{\rm pot}(\ell_1-\ell_2), \\
 \hat{T} &= -t \sum_\ell \ket{\ell+1} \bra{\ell} + \hc. ~.
\end{flalign*}
Note that the $k_y$-dependence of $ \Delta_{\rm S}(k_y) $ and $J_{\rm ex}(k_y)$ enters because the optimized wavefunction $\ket{\phi^{(1)}(k_y)}$ depends on $k_y$ (we dropped the explicit dependence in the above expressions for clarity).

\begin{figure}[t!]
\centering
\epsfig{file=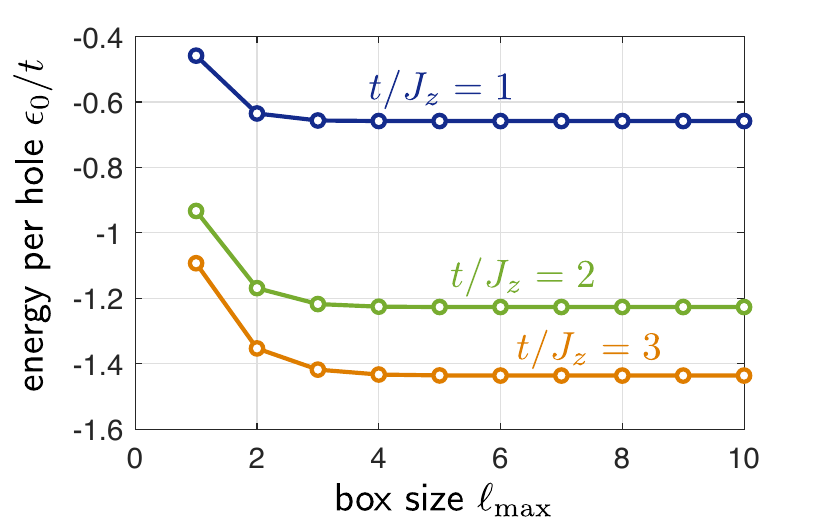, width=0.4\textwidth} $\qquad$
\caption{To estimate the effective repulsive potential between stripes, we calculate the energy per hole $\epsilon_0$ of one stripe in a box extending from $- \ell_{\rm max} ... \ell_{\rm max}$.} 
\label{figStringPot}
\end{figure}

\emph{Lattice gas model.--}
When more than one hole per chain is considered we expect stripes to repel each other weakly along $x$-direction: Chargons are mutually hard-core particles, which leads to an additional localization energy cost when two chargons come close to each other. To estimate the strength of the repulsive stripe-potential, we calculate the ground state energy per hole for a stripe in a box from $x=-\ell_{\rm max} ... \ell_{\rm max}$, see Fig.~\ref{figStringPot}. On one hand this gives an estimate of the energy cost to localize stripes within a region $\pm \ell_{\rm max}$. On the other hand our results show for the dilute regime (no stripe interactions) that our calculations quickly converge with $\ell_{\rm max}$.

%%%%%%%%%%%%%%%%%%%%%%%%%%%%%%%%%%%%%%%%%%%%%%%%%%%%%
%%%%%%%%%%%%%%%%%%%%%%%%%%%%%%%%%%%%%%%%%%%%%%%%%%%%%

The classical energy of multiple stripes at positions $x_j$, where $j=1... L_x n_{\rm h}$, can be modeled by
\begin{equation}
 \mathscr{H} = \sum_{i<j=1}^{n_{\rm h} L_x} V_{\rm S}(|x_i - x_j|);
 \label{eqDefHlatticeGas}
\end{equation}
this corresponds to the Hamiltonian of a classical 1D Ising model (lattice gas model). We estimate the effective potential $V_{\rm S}(d) \approx \epsilon_0(d)  - \epsilon_0(\infty)$ from the energy per hole for one string in a box, see Fig.~\ref{figEllS}. For the smallest distances $V_{\rm S}(1) \approx 0.2 t$ is a small fraction of $t$. Since the solution $\phi^{(0)}_\ell$ asymptotically behaves like the Airy function ${\rm Ai}(\ell)$ (due to the linear string potential Eq.~\eqref{eqVellDef} at large distances), this potential decays super-exponentially: 
\begin{equation}
V_{\rm S}(\ell) \simeq \exp [ - C \ell^{2/3}].
\label{eqVSdecay}
\end{equation}

The ground state of the model Eq.~\eqref{eqDefHlatticeGas} has long-range order (charge density wave, CDW) along $x$. 

\emph{Non-zero temperatures: phonons.--}
Next we discuss the stability of the stripes forming along $y$ when thermal fluctuations are included. The phonon excitation spectrum of a single stripe, see Eq.~\eqref{eqOmegaPhonon}, is gapped: $\Delta = \Delta_{\rm S}(0) - 2 J_{\rm ex}(0)$. Therefore we expect that stripes are robust as long as $k_B T \lesssim \Delta$. In the limit of small doping $n_{\rm h} \to 0$ we  show below, however, that the meson gas has a smaller free energy. 

To estimate the free energy per hole in the stripe phase, $F_{\rm S} / N_{\rm h}$, we neglect phonon ($\b_{k_y}$) non-linearities and describe fluctuations on top of the stripe phase by the free phonon Hamiltonian
\begin{equation}
 \H_{\rm ph} = \sum_{k_y} \omega_{\rm ph}(k_y) ~ \bd_{k_y} \b_{k_y}.
 \label{eqHeffPhonon}
\end{equation}
This leads to an ideal-Bose gas contribution to the free energy of
\begin{equation}
 \frac{F_{\rm S, ph} }{ n_{\rm h} L_x}= k_B T  \frac{L_y}{2 \pi} \int_{-\pi}^\pi dk_y~ \log \l 1 - e^{- \omega_{\rm ph}(k_y) / k_B T} \r
 \label{eqFstrpPh}
\end{equation}
per stripe.

There is a second contribution $F_{{\rm S, cl}}$ to the free energy, related to classical fluctuations of the locations $x_j$ of the stripes, $j=1... n_{\rm h} L_x$. This contribution to the free energy vanishes in the thermodynamic limit $L_{x}, L_y \to \infty$. To see this, we give a simple estimate by focusing on the entropic contribution: $F_{{\rm S, cl}} \approx - k_B T  \log(M_{\rm S})$ where $M_{\rm S}$ denotes the number of microscopic stripe configurations. In the dilute regime when $n_{\rm h} \ll 1$ we expect that the energetic contribution beyond the ground state is negligible since the stripe interaction potential decays super-exponentially, see Eq.~\eqref{eqVSdecay}. Assuming further that the width of each stripe, $\sigma_{\rm S} = [\bra{\phi^{(0)}} \hat{\ell}^2 \ket{\phi^{(0)}} ]^{1/2}$, is small compared to the average distance, $\sigma_{\rm S} \ll 1/n_{\rm h}$, we obtain:
\begin{equation}
 \frac{F_{{\rm S,cl}}}{n_{\rm h} L_x} \approx - k_B T \left[ 1 - \log \l \sigma_{\rm S} n_{\rm h} \r \right].
\end{equation}
This result is independent of $L_y$, and thus sub-dominant compared to $F_{\rm S, ph}$.

Combining our results, we obtain
\begin{multline}
\frac{F_{\rm S}}{n_{\rm h} L_x L_y}  =  \epsilon_0^{\rm S} + k_B T \int_{- \pi}^\pi \frac{dk_y}{2 \pi}  \log \l 1 - e^{- \frac{\omega_{\rm ph}(k_y) }{ k_B T}} \r \\
 + \mathcal{O}(1/L_y) 
 \label{eqFSresult}
\end{multline}
where $\epsilon_0^{\rm S}$ is the ground state energy per hole relative to the undoped N\'eel state, see Eq.~\eqref{eqE0SperHole}. Eq.~\eqref{eqFSresult} is valid when the average distance of stripes is well beyond their width, $n_{\rm h} \ll \sigma_{\rm S}$: this justifies neglecting interactions between stripes.

\emph{Non-zero temperatures: Mean-field.--}
At higher temperatures $k_B T \gtrsim \Delta_{\rm S} - 2 J_{\rm ex}$, the inclusion of only the lowest phonon band \eqref{eqHeffPhonon} is no longer sufficient. To describe thermal fluctuations in this regime, we generalize the mean-field ansatz from Eq.~\eqref{eqDefMFstripes} to a product of density matrices:
\begin{equation}
 	\hat{\rho}_{\rm S, MF} =   \prod_{y=1}^{L_y} \hat{\rho}^{(0)}_{\rm MF}  =  \prod_{y=1}^{L_y} \left[  e^{- \H_{\rm MF} / k_B T } / Z_{\rm MF} \right],
	\label{eqDefMFstripesT}
\end{equation}
where $\H_{\rm MF}$ is the mean-field Hamiltonian from Eq.~\eqref{eqDefHMFstripes} with a temperature dependent potential $V_{\rm eff}(\ell ; T)$. 

%%%%%%%%%%%%%%%%%%%%%%%%%%%%%%%%%%%%%%%%%%%%%%%%%%%%%
\begin{figure}[t!]
\centering
\epsfig{file=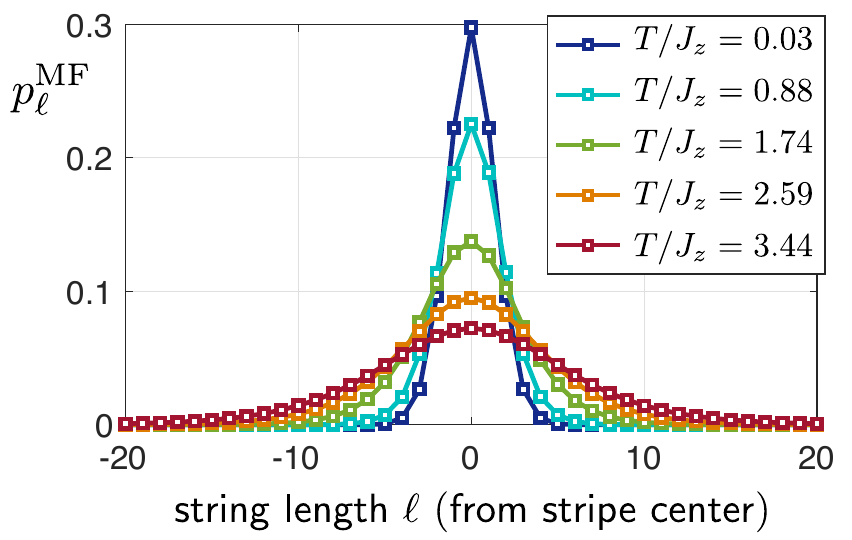, width=0.4\textwidth} $\quad$
\caption{String length distribution $p_\ell^{\rm MF}$ of stripes at non-zero temperature, measured from the stripe center. We used $\ell_{\rm max}=30$ in the calculation.} 
\label{figEllStherm}
\end{figure}
%%%%%%%%%%%%%%%%%%%%%%%%%%%%%%%%%%%%%%%%%%%%%%%%%%%%%

As for the ground state, one can either solve for the mean-field state $\hat{\rho}_{\rm S, MF}$ by determining the mean-field potential self-consistently:
\begin{equation}
 V_{\rm eff}(\ell ; T) = 2 \sum_r \bra{r} \hat{\rho}^{(0)}_{\rm MF} \ket{r} ~ V_{\rm pot}(r - \ell).
\end{equation}  
Alternatively, we choose to treat $V_{\rm eff}(\ell)$ as a set of free parameters which determine $\hat{\rho}^{(0)}$ and can be optimized variationally. To this end we minimize the free energy per hole (relative to the energy of the undoped N\'eel state):
\begin{multline}
 \frac{F_{\rm S}(T)}{n_{\rm h} L_x L_y} =  - k_B T \log \l \sum_n e^{- E_n^{\rm MF} / k_B T} \r + \\
+ \sum_\ell p_\ell^{\rm MF} \left[ \sum_r V_{\rm pot}(\ell - r) p_r^{\rm MF} - V_{\rm eff}(\ell ; T) \right].
\end{multline}
Here $E_n^{\rm MF}$ denote the eigenenergies of the mean-field Hamiltonian $\H_{\rm MF} + L_x J_z/2$ relative to the undoped N\'eel state and the mean-field string-length distribution is given by
\begin{equation}
 p_\ell^{\rm MF} = {\rm tr}  \left[ \hat{\rho}^{(0)}_{\rm MF} \ket{\ell} \bra{\ell}  \right].
\end{equation}
This string length distribution is shown for the variationally optimized mean-field state in Fig.~\ref{figEllStherm}.

%%%%%%%%%%%%%%%%
\subsubsection{Mean-field theory of the meson gas}
%%%%%%%%%%%%%%%%
Next we describe individual mesons, composed of two chargons, in the meson gas regime. We start by considering a single pair of two holes, and extend our results to finite densities afterwards.

\emph{Two individual chargons.--}
Our starting point is the undoped N\'eel state, for which $\tau^x_\ij \equiv 1$ everywhere. Next we add two chargons in the same chain, which are connected by a \Zt electric string along which $\tau^x_\ij = - 1$. This leads to a linear potential depending on the distance $\ell$ of the two chargons,
\begin{equation}
 V_{\rm cc}(\ell) = m  - \frac{J_z}{4} \delta_{|\ell |,1}+  \frac{dE}{d \ell} \l |\ell| - 1 \r,
 \label{eqDefVcc}
\end{equation}
where $dE/d\ell = J_z$ is the linear string tension and $m=2 J_z$ a zero-point energy shift. Note that the string length $\ell \neq 0$ cannot become zero since chargons are mutually hard-core.  

Since the motion of chargons is restricted to one dimension, along $x$, we can treat them as distinguishable. Transforming into the co-moving frame with the left chargon yields the following effective Hamiltonian in the relative coordinate $\ell>0$:
\begin{equation}
 	\H_{\rm cc}(k_x) =  \sum_{\ell > 0} \biggl[ \bigl(  t_{\rm cc}(k_x) \ket{\ell+1} \bra{\ell} +\hc \bigr) + \ket{\ell}\bra{\ell} V_{\rm cc}(\ell) \biggr].
	\label{eqDefHcc}
\end{equation}
Here energies are measured relative to the undoped N\'eel state; $k_x$ denotes the total conserved momentum of the meson and the effective tunnel coupling is given by
\begin{equation}
 t_{\rm cc}(k_x) = t \sqrt{2 \l 1 + \cos k_x \r}.
\end{equation}

Using exact diagonalization, the spectrum $\epsilon_{\rm cc}^{n}(k_x)$ of \eqref{eqDefHcc} can be easily computed. This yields the ground state, $n=0$, and allows to include thermal fluctuations by including higher vibrational excitations $n\geq 1$ with their respective Boltzmann weights. 

\emph{Finite doping.--}
To calculate the free energy per hole at non-zero doping, $n_{\rm h} > 0$, we treat the chargon-chargon mesons as point-like hard-core bosons $\hat{d}_{k_x,n}^{(\dagger)}$. This is justified if the average string length $\langle \ell_{\rm cc} \rangle = \langle \sum_\ell \ell ~ \ket{\ell} \bra{\ell} \rangle$ is small compared to the average meson distance: $\langle \ell_{\rm cc} \rangle \ll 2 / n_{\rm h}$. By introducing Jordan-Wigner strings, the mesons $\hat{d}_{k_x,n}^{(\dagger)} \to \hat{f}_{k_x,n}^{(\dagger)}$ are fermionized and we use the following free fermion Hamiltonian per chain:
\begin{equation}
\H_{\rm cc, eff} = \sum_{n \geq 0} \int_{- \pi}^\pi dk_x ~ \hat{f}^\dagger_{k_x,n} \hat{f}_{k_x,n} ~ \epsilon_{\rm cc}^n (k_x);
\end{equation}
$n$ denotes the internal vibrational states of the mesons. 

Neglecting interactions between mesons in different chains for now, and working in the grand-canonical ensemble, we obtain the following free energy per hole:
\begin{multline}
\frac{F^{(0)}_{\rm cc}}{n_{\rm h} L_x L_y} = \frac{\mu^{(0)}_{\rm cc}}{2} - k_B T \frac{n_{\rm h}^{-1}}{2 \pi} \int_{-\pi}^\pi dk_x~ \sum_n \\
 \times \log \left[  1 + \exp \l - \frac{\epsilon_{\rm cc}^n(k_x) - \mu^{(0)}_{\rm cc}}{k_B T} \r \right].
 \label{eqDefF0cc}
\end{multline}
As usual, we define $F^{(0)}_{\rm cc}$ relative to the undoped state. The meson chemical potential $\mu_{\rm cc}^{(0)}$ needs to be fixed independently by solving
\begin{equation}
 n_{\rm h} = \frac{1}{\pi} \int_{-\pi}^\pi dk_x  \sum_n \left[ 1 + \exp \l \frac{\epsilon_{\rm cc}^n(k_x) - \mu^{(0)}_{\rm cc}}{k_B T} \r  \right]^{-1}.
 \label{eqDefmucc}
\end{equation}

\emph{Mean-field interaction effects.--}
On a mean-field level, interactions between mesons from different chains can be easily taken into account: They lead to a reduction of the linear string tension $dE/d\ell$ and the zero-point energy shift $m$. To calculate the renormalized meson potential parameters we start from the mean-field Hamiltonian:
\begin{multline}
\H_{\rm MF}^{\rm cc} = \H_t + \frac{J_z}{2} \sum_{\vec{j}} \hat{n}^h_{\vec{j}} - \frac{J_z}{4} \sum_{\ij_x} \hat{n}^h_{\vec{i}} \hat{n}^h_{\vec{j}} \\
- \frac{J_z}{4} (1-n_{\rm h}) \langle \hat{\tau}^x \rangle_{\rm MF} \sum_{\ij_x}  \hat{\tau}^x_{\ij_x} (2-n^h_{\vec{i}} - n^h_{\vec{j}}),
\label{eqHMFcc}
\end{multline}
where $\H_t$ denotes the hopping part of chargons; note a factor of $2$ in the second term which accounts for interactions with both neighboring chains in $y$-direction. 

In the meson mean-field phase, the average \Zt electric field is related to the average meson string length:
\begin{equation}
  \langle \hat{\tau}^x \rangle_{\rm MF} = 1 - n_{\rm h} \langle \ell_{\rm cc} \rangle.
\end{equation}
Using this result we read off from the mean-field chargon Hamiltonian, Eq.~\eqref{eqHMFcc}: 
\begin{flalign}
\frac{dE}{d \ell} |_{\rm MF} &= J_z  (1-n_{\rm h}) \l 1 - n_{\rm h} \langle \ell_{\rm cc} \rangle \r, \\
m  |_{\rm MF} &= J_z \left[ 1 + \l 1 - n_{\rm h} \langle \ell_{\rm cc} \rangle \r \l 1 - n_{\rm h} \r \right].
\end{flalign}
Note that the resulting chargon potential $V^{\rm MF}_{\rm cc}(\ell)$ from Eq.~\eqref{eqDefVcc} counts the mean-field energy Eq.~\eqref{eqHMFcc} of mesons relative to the energy $E_0^{\rm MF}$ of an undoped chain described by the same mean-field Hamiltonian (the latter including doping in the neighboring chains). 

To evaluate the resulting free energy, $F_{\rm cc} = \langle \H \rangle_{\rm MF} - T S_{\rm MF}$, we note that the mean-field Hamiltonian Eq.~\eqref{eqHMFcc} yields $F_{\rm cc}^{\rm MF} = \langle \H_{\rm MF}^{\rm cc} \rangle_{\rm MF} - T S_{\rm MF}$. This gives
\begin{equation}
 F_{\rm cc} = F_{\rm cc}^{\rm MF} +  \langle \H \rangle_{\rm MF} - \langle \H_{\rm MF}^{\rm cc} \rangle_{\rm MF} .
\end{equation}
Since the true and mean-field Hamiltonians only differ by their potential energy terms,
\begin{equation}
 \langle \H \rangle_{\rm MF} - \langle \H_{\rm MF}^{\rm cc} \rangle_{\rm MF} = L_x L_y \frac{J_z}{4} (1-n_{\rm h})^2  \l 1 - n_{\rm h} \langle \ell_{\rm cc} \rangle \r^2,
\end{equation}
we can write the free energy of the mean-field state -- now measured relative to the energy $E_0^{\rm N} = - L_x L_y J_z/2$ of the undoped N\'eel state -- as:
\begin{multline}
  F_{\rm cc} = L_y \l f_{\rm cc}^{{\rm MF},0} + \varepsilon_0^{\rm MF} + L_x J_z/2  \r + \\
  +  L_x L_y \frac{J_z}{4} (1-n_{\rm h})^2  \l 1 - n_{\rm h} \langle \ell_{\rm cc} \rangle \r^2.
  \label{eqFccDerivation}
\end{multline}
Here $f_{\rm cc}^{{\rm MF},0}$ is the mean-field meson free energy per chain (described by the chargon potential Eq.~\eqref{eqDefVcc} with mean-field parameters) measured relative to the energy 
\begin{equation}
\varepsilon_0^{\rm MF} = - L_x J_z  \left[ \frac{1}{4} +  (1-n_{\rm h}) \l 1 - n_{\rm h} \langle \ell_{\rm cc} \rangle \r \frac{1}{2} \right] 
\end{equation}
of an undoped chain in the mean-field Hamiltonian. Simplifying all terms in Eq.~\eqref{eqFccDerivation} yields
\begin{equation}
 F_{\rm cc} = F_{\rm cc}^{{\rm MF},0} + L_x L_y \frac{J_z}{4} n_{\rm h}^2 \biggl( 1 + \langle \ell_{\rm cc} \rangle - n_{\rm h} \langle \ell_{\rm cc} \rangle \biggr)^2.
\label{eqFccRes}
\end{equation}

The mean-field free energy obtained per meson, $F_{\rm cc}^{{\rm MF},0} / (n_{\rm h} L_x L_y) \equiv f_{\rm cc}^{{\rm MF},0} / (n_{\rm h} L_x)$, can be calculated as in Eqs.~\eqref{eqDefF0cc}, \eqref{eqDefmucc} but using the renormalized mean-field chargon spectrum $\epsilon_{\rm cc}^n(k_x) \to \epsilon_{\rm cc}^{{\rm MF}, n}(k_x)$: 
\begin{multline}
\frac{F^{{\rm MF},0}_{\rm cc}}{n_{\rm h} L_x L_y} = \frac{\mu^{\rm MF}_{\rm cc}}{2} - k_B T \frac{n_{\rm h}^{-1}}{2 \pi} \int_{-\pi}^\pi dk_x~ \sum_n \\
 \times \log \left[  1 + \exp \l - \frac{\epsilon_{\rm cc}^{{\rm MF}, n}(k_x) - \mu^{\rm MF}_{\rm cc}}{k_B T} \r \right],
 \label{eqDefF0ccMF}
\end{multline}
and:
\begin{equation}
 n_{\rm h} = \frac{1}{\pi} \int_{-\pi}^\pi dk_x  \sum_n \left[ 1 + \exp \l \frac{\epsilon_{\rm cc}^{{\rm MF},n}(k_x) - \mu^{\rm MF}_{\rm cc}}{k_B T} \r  \right]^{-1}.
 \label{eqDefmuccMF}
\end{equation}

%%%%%%%%%%%%%%%%
\subsubsection{Mean-field theory of the chargon gas}
%%%%%%%%%%%%%%%%
For the mean-field theory of the chargon gas we assume that chargons form independent free Fermi gases in each chain:
\begin{equation}
  \ket{\Psi_{\rm ch}} =  \prod_{y=1}^{L_y}  \ket{\phi_{\rm FS}}_y.
\end{equation}
A corresponding product ansatz of free-fermion density matrices can be made for non-zero temperatures. For product states as in Eq.~\eqref{eqDefPsiCh}, the mean-field Hamiltonian becomes
\begin{multline}
 \H_{\rm MF, ch} = - L_x L_y \frac{J_z}{4} - t \sum_{\ij_x} \l \hd_{\vec{i}}  ~\hat{\tau}^z_{\ij_x} \h_{\vec{j}} + \hc \r +\\
 + \frac{J_z}{2} \sum_{\vec{j}} \hat{n}^h_{\vec{j}} - \frac{J_z}{4} \sum_{\ij_x} \hat{n}^h_{\vec{i}} \hat{n}^h_{\vec{j}},
 \label{eqHMFchgas}
\end{multline}
where inter-chain couplings are absent because we assume that $\langle \hat{\tau}^x_{\ij_x} \rangle_{\rm MF} = 0$ in the chargon gas phase. 

The Hamiltonian \eqref{eqHMFchgas} can be diagonalized by introducing the following fermionic operators \cite{Prosko2017}
\begin{equation}
 \hat{f}_{\vec{r}} = \prod_{\ij_x \geq \vec{r}} \l \hat{\tau}^z_{\ij_x} \r \h_{\vec{r}},
\end{equation}
where the product is over all links in the same chain and right from site $\vec{r}$. Using $\hat{n}^f_{\vec{j}} = \hat{f}^\dagger_{\vec{j}} \hat{f}_{\vec{j}} = \hat{n}^h_{\vec{j}}$ one gets
\begin{equation}
  \H_{\rm MF, ch} = - L_x L_y \frac{J_z}{4} + \sum_{\vec{k}}  \varepsilon_{\rm c}(k_x) \hat{f}^\dagger_{\vec{k}} \hat{f}_{\vec{k}} - \frac{J_z}{4} \sum_{\ij_x} \hat{n}^f_{\vec{i}} \hat{n}^f_{\vec{j}},
\end{equation}
with the free chargon dispersion
\begin{equation}
 \varepsilon_{\rm c}(k_x) = - 2 t \cos (k_x) + \frac{J_z}{2}.
\end{equation}

Ignoring the weak nearest-neighbor attraction $-J_z / 4$ leads us to the variational ansatz of a free Fermi gas:
\begin{equation}
 \ket{\Psi_{\rm ch}} = \prod_{y=1}^{L_y}  \prod_{|k_x|<k_{\rm F}} \hat{f}^\dagger_{k_x,y} \ket{0},
 \label{eqDefPsiCh}
\end{equation}
with $k_{\rm F} = \pi n_{\rm h}$ denoting the Fermi momentum, and similarly for density matrices at non-zero temperatures. This leads to the following free energy per hole,
\begin{multline}
 \frac{F_{\rm ch}}{n_{\rm h} L_x L_y} = \frac{1}{n_{\rm h}} \frac{J_z}{4} - n_{\rm h} \frac{J_z}{4} g_1^{(2)}(n_{\rm h},T) + \mu_{\rm c} \\
 - k_B T \frac{n_{\rm h}^{-1}}{2 \pi} \int_{- \pi}^\pi dk_x ~  \log \left[  1 + \exp \l - \frac{\varepsilon_{\rm c}(k_x) - \mu_{\rm c}}{k_B T} \r \right]
 \label{eqFchRes}
\end{multline}
measured relative to the undoped N\'eel state. Here we introduced the nearest-neighbor chargon correlator,
\begin{multline}
 g^{(2)}_1(n_{\rm h},T) = \frac{\langle \hat{n}^f_x \hat{n}^f_{x+1} \rangle}{n_{\rm h}^2} = 1 - \frac{1}{n_{\rm h}^2 } \times \\
 \times \l \int_{-\pi}^\pi \frac{dk_x}{2 \pi} ~ \cos (k_x) \left[ 1 + e^{( \varepsilon_{\rm c}(k_x) - \mu_{\rm c} ) / k_B T} \right]^{-1} \r^2,
\end{multline}
and the chargon chemical potential $\mu_{\rm c}$ is determined by
\begin{equation}
 n_{\rm h} = \frac{1}{2 \pi} \int_{- \pi}^\pi dk_x ~ \left[ 1 + e^{( \varepsilon_{\rm c}(k_x) - \mu_{\rm c} ) / k_B T} \right]^{-1}
\end{equation}

%%%%%%%%%%%%%%%%
\subsubsection{Results}
\label{SecSMresults}
%%%%%%%%%%%%%%%%
Now we compare the results collected in the last few subsections. We start at zero temperature, where the free energy $F = E$ reduces to the (variational) ground state energy. In the following we calculate energies per hole, where $N_{\rm h} = L_x L_y n_{\rm h}$ denotes the total number of holes in the system. All energies are measured relative to the undoped 2D N\'eel state.

\emph{Mean-field phase diagram.--}
The mean-field phase diagram in Fig.~\ref{figOverview} (b) of the main text indicates which of the three considered phases has the lowest free energy. In the calculation of the meson gas free energy, we included meson-meson interactions on a mean-field level, using Eq.~\eqref{eqDefF0ccMF}. We used the finite-temperature Gutzwiller ansatz, Eq.~\eqref{eqDefMFstripesT}, for the stripes, but found essentially identical results by using the one-phonon approximation Eq.~\eqref{eqFstrpPh}.

We only calculated the meson gas free energy in a regime where the average string length $\langle \ell_{cc} \rangle < 1 / n_{\rm h}$, which is required to assume point-like mesons in our calculation. Likewise, we only calculated the stripe free energy in regimes where $\langle \hat{\ell} \rangle < 0.5 / n_{\rm h}$, beyond which stripe-stripe interactions must be included. This leads to the cusps and re-entrant behaviors we find in the mean-field phase diagram. We do not expect these to be physically meaningful.

\emph{Zero temperature.--}
Neglecting interactions between stripes along $x$, we find for the ground state energy per hole in the stripe phase, see Eq.~\eqref{eqFSresult}:
\begin{equation}
 \frac{F_{\rm S}}{N_{\rm h}} = \epsilon_0^{\rm S}.
 \label{eqFSsumm}
\end{equation}
For the meson gas we obtain a similar result, see Eq.~\eqref{eqFccRes}:
\begin{equation}
 \frac{F_{\rm cc}}{N_{\rm h}} = \frac{1}{2} \epsilon_{\rm cc}^{n=0}(k_x=0) + \mathcal{O}(n_{\rm h}),
  \label{eqFccsumm}
\end{equation}
where we ignored mean-field effects yielding $\mathcal{O}(n_{\rm h})$ corrections. For the chargon gas Eq.~\eqref{eqFchRes} gives
\begin{equation}
 \frac{F_{\rm ch}}{N_{\rm h}} = \frac{J_z}{4} \times \frac{1}{n_{\rm h}} + \frac{J_z}{2} - 2 t +  \mathcal{O}(n_{\rm h}).
  \label{eqFchsumm}
\end{equation}
The last term captures finite-density corrections $\mathcal{O}(n_{\rm h})$.

Comparing Eqs.~\eqref{eqFSsumm} - \eqref{eqFchsumm} we note that the energy per hole of the chargon gas diverges when $n_{\rm h} \to 0$. This is understood by noting that the chargon gas has vanishing AFM correlations along $y$ at any non-zero doping $n_{\rm h}>0$, and thus does not approach the 2D N\'eel state when $n_{\rm h} \to 0$. This leads to a total energy cost $\mathcal{O}(L_x L_y)$ at $n_{\rm h}=0$, which translates into a diverging energy \emph{per hole} in this regime. We conclude that the chargon gas can only exist beyond a critical doping $n_{\rm h}^{\rm ch} > 0$; below $n_{\rm h}^{\rm ch}$ only the stripe and meson gas compete. 

To compare stripes and the meson gas, we first compare their asymptotic ground state energies per hole at infinitesimal doping: $\epsilon_0^{\rm S}$ and $ \nicefrac{1}{2} ~ \epsilon_{\rm cc}^{n=0}(k_x=0)$. We find that the stripe consistently has a lower energy per hole than the two-chargon meson. 

\emph{Low-temperature, low doping: meson gas vs. stripes.--}
To understand the behavior at small but non-zero temperatures, we expand the stripe free energy in $k_B T$. Further, since we also focus on small doping $n_{\rm h} \ll 1$, we only compare stripe and meson gas phases relevant to this regime in the following. 

For the stripe phase we can capture thermal fluctuations using low-energy phonon excitations, see Eq.~\eqref{eqFSresult}. The non-zero phonon gap, $\Delta = \Delta_{\rm S}(0) - 2 J_{\rm ex}(0)$, leads to a finite activation energy and corresponding exponential suppression of thermal fluctuations for stripes:
\begin{equation}
 \frac{F_{\rm S}}{N_{\rm h}} \simeq \epsilon_0^{\rm S} - k_B T e^{- \frac{\Delta}{k_B T}}, \quad k_B T \ll \Delta.
 \label{FSlowTexp}
\end{equation}

For the meson gas in the same regime, we can neglect vibrational excitations with $\epsilon_{\rm cc}^{n>0} > \epsilon_{\rm cc}^{n=0}$ and consider a free Fermi gas (note that mesons are hard-core particles moving in 1D chains) in the lowest band $\epsilon_{\rm cc}^0(k_x)$. Expanding $\epsilon_{\rm cc}^0(k_x) = \epsilon_{\rm cc}^0 + k_x^2 / 2 m_{\rm cc}$ and working in the low doping regime, we obtain a classical gas with free energy per hole:
\begin{equation}
 \frac{F_{\rm cc}}{N_{\rm h}} \simeq \frac{1}{2} \epsilon_{\rm cc}^0  - k_B T  \frac{1}{2}  \Bigl( 1 - \log( n_{\rm h} / 2) \Bigr).
  \label{FcclowTexp}
\end{equation}

Comparing Eqs.~\eqref{FSlowTexp}, \eqref{FcclowTexp} yields the transition temperature from stripes to the meson gas when $n_{\rm h} \ll 1$:
\begin{equation}
 k_B T_c = \frac{\epsilon_{\rm cc}^0  - 2 \epsilon_0^{\rm S}}{1 - \log( n_{\rm h} /2 )}.
\end{equation}

\emph{High-temperature expansion: chargon vs. meson gas.--}
We have already seen in the previous paragraph that stripes have too low entropy at high temperatures. Therefore we only consider the gaseous phases in the following.

At high temperatures, the  free energy per hole of the chargon gas, Eq.~\eqref{eqFchRes}, becomes
\begin{equation}
 \frac{F_{\rm ch}}{n_{\rm h} L_x L_y} = \frac{1}{n_{\rm h}} \frac{J_z}{4} - n_{\rm h} \frac{J_z}{4} + \frac{J_z}{2} + k_B T \left[ \log(n_{\rm h}) - 1 \right].
 \label{eqFchHighTLowNh}
\end{equation}
This term is still dominated by the first divergent contribution at low loping $n_{\rm h} \to 0$. 

To estimate the free energy per hole at high temperatures $k_B T \gg t$ and low doping $n_{\rm h} \ll 1 / \langle \ell_{\rm cc} \rangle$ for the meson gas phase, we note that the meson dispersion can be approximated by
\begin{equation}
 \epsilon_{\rm cc}^n(k_x) \approx n J_z
\end{equation}
when $\epsilon_{\rm cc}^n \gg t$. In this regime, the eigenstates correspond to Wannier-Stark states in the linear potential $V_{\rm cc}(\ell) \approx J_z \ell$. The corresponding free energy per hole is purely classical and becomes:
\begin{equation}
  \frac{F_{\rm cc}}{n_{\rm h} L_x L_y} \approx - \frac{1}{2} k_B T \log \l \frac{k_B T}{J_z} \r.
  \label{eqFccHighTLowNh}
\end{equation}
The average string length can be estimated in a similar manner,
\begin{equation}
 \langle \ell_{\rm cc} \rangle \approx \frac{k_B T }{J_z},
\end{equation}
when $k_B T \gg t$.

Comparison of Eqs.~\eqref{eqFchHighTLowNh}, \eqref{eqFchHighTLowNh} shows that the meson gas is thermodynamically favored in the low doping, high temperature regime.

%%%%%%%%%%%%%%%%%%%%%%%%%%%%%%%%%%%%%%%%
%\bibliography{/Users/fgrusdt/Documents/Library/dataBase_JabRef2.bib}
%\bibliography{/Users/fabianbohrdt/Documents/Library/dataBase_JabRef2.bib}
\bibliographystyle{unsrt}

\end{document}